\documentclass[usenatbib,useAMS]{mn2e}
\usepackage{times}
\usepackage{epsf}
\usepackage{amssymb}

\newcommand{\bv}{{\bf v}}

\newcommand{\calR}{{\cal R}}
\newcommand{\calL}{{\cal L}}
\newcommand{\tilL}{\tilde{\calL}}
\newcommand{\tilT}{\tilde{T}}
\newcommand{\tilrho}{\tilde{\rho}}
\newcommand{\tilp}{\tilde{p}}
\renewcommand{\d}{{\rm d}}
\newcommand{\ghat}{{\hat{\textnormal{\mathversion{bold}$g$}}}}

\title[Dynamics of curved interfaces]{Evaporation and condensation of HI clouds
in thermally bistable interstellar media: semi-analytic description of
isobaric dynamics of curved interfaces}

\author[M. Nagashima, H. Koyama and S. Inutsuka]{
Masahiro Nagashima,$^{1}$\thanks{E-mail:
masa@scphys.kyoto-u.ac.jp (MN)}
Hiroshi Koyama,$^{2}$
and 
Shu-ichiro Inutsuka$^{1}$\\
$^{1}$Department of Physics, Graduate School of Science,
Kyoto University, Sakyo-ku, Kyoto 606-8502, Japan\\
$^{2}$Department of Earth and Planetary Science, Kobe
University, Kobe 657-8501, Japan}

\begin{document}
\maketitle

\begin{abstract}
We analyse the evaporation and condensation of spherical and cylindrical
HI clouds of the cold neutral medium surrounded by the warm neutral
medium.  Because the interstellar medium including those two phases is
well described as a thermally bistable fluid, it is useful to apply
pattern formation theories to the dynamics of the interface between the
two phases.  Assuming isobaric evolution of fluids and a simple
polynomial form of the heat-loss function, we show the curvature effects
of the interface.  We find that approximate solutions for spherical
clouds are in good agreement with numerically obtained solutions.  We
extend our analysis to general curved interfaces taking into account the
curvature effects explicitly.  We find that the curvature effects always
stabilise curved interfaces under assumptions such as isobaric evolution
we adopt in this {\it Letter}.
\end{abstract}


\section{Introduction}
It is widely known that the interstellar medium (ISM) is well described
as a thermally bistable fluid owing to radiative cooling and heating due
to external radiation fields and cosmic rays \citep{fgh69, wolfire03}.
Two stable phases are called the warm neutral medium (WNM) with
temperature $T\sim 10^{4}$ K and the cold neutral medium (CNM) with
$T\sim 10^{1-2}$ K, respectively.  Gas in an unstable phase with
temperature between the WNM and CNM is decomposed into the two stable
phases via thermal instability \citep{f65, b86}.  These stable phases
can be connected through interfaces in pressure equilibrium.  It is
important for understanding the behaviour of the ISM such as the
interstellar turbulence \citep{ki02, ki04, kn02a, kn02b, ah05} and the
evolution of galaxies \citep[e.g.][]{mo77} to clarify the dynamics of
the interface or {\it front}, which corresponds to the evaporation and
condensation of low temperature HI clouds.

\citet{zp69} and \citet{pb70} considered isobaric and steady
phase-change in the bistable fluid assuming plane-parallel geometry.  In
this case, because the mass flux across the interface conserves, the
velocity of fluids can be estimated as an eigenvalue of the energy
equation.  Then they pointed out the relationship between the motion of
fronts, which determines the rate of evaporation or condensation of
clouds, and external pressure, and the existence of {\it saturation}
pressure at which a static front can exist.  After that, many authors
have investigated the thermally bistable flow \citep[e.g.][]{fs93, hp99,
svg02}.  \citet{ers91} and \citet[][hereafter ERS92]{ers92} treated with
fluid equations in a more sophisticated manner in plane-parallel
geometry.  In the latter work, they formulated equations in the
Lagrangian coordinate.  Combined with the isobaric assumption, they
derived a second-order ordinary differential equation in a steady state
from the energy equation.  Moreover, they discussed the behaviour of
steady solutions from a pattern-theoretical point of view.

It is challenging to extend those analyses to higher dimensions while it
is much useful for applying to realistic situations.  \citet{gl73}
numerically computed isobaric flows in three dimensional spherical
geometry for the first time.  They pointed out the existence of a
critical size of clouds to avoid evaporation.  Based on ERS92,
\citet{sr94} argued the case of higher dimensions.  However, it is
difficult to extend the Lagrangian formulation in plane-parallel
geometry provided by ERS92 to higher dimensional geometry in a
straightforward manner.  Therefore assuming a model equation similar to
the Ginzburg-Landau (GL) equation, they derived the speed of frontal
motion.  Surprisingly, they have showed that the dependence of the
frontal speed on time, or radii of clouds, is dependent on the dimension
of geometry.  Besides they have claimed that their conclusion is
supported by numerical simulations.  It should be noted that
\citet{ams95} and \citet{m96} also discussed the front curvature effects
in a confined plasma, in which the boundary conditions are different
from ours.  Thereby their methods are inapplicable to the ISM in a
straightforward way.

Thus, the purpose of this paper is to reanalyse the frontal motion in
$d$-dimensional spherically symmetric geometry, aided by pattern
formation theories \citep[e.g.][]{b94}.  This paper is outlined as
follows.  In Section 2 we briefly review ERS92 to prepare the analysis
in higher dimensional geometry.  In Section 3 we show a systematic
procedure to derive the curvature effects.  In Section 4 we discuss the
dynamics of general curved fronts.  In Section 5 we provide conclusions
and discussion.

\section{Lagrangian description in plane-parallel geometry}
In the following, we assume isobaric evolution in which pressure is
uniform over the whole system because the fluid motion is much slower
than the sound speed.  The basic fluid equations under the assumption of
isobaric evolution are written as
\begin{eqnarray}
 &&\frac{\partial\rho}{\partial t}+\nabla\cdot\rho\bv=0,\\
 &&\frac{\gamma}{\gamma-1}\frac{\calR}{\mu}\rho\frac{\d T}{\d t}+\rho\calL-\nabla\cdot\kappa(T)\nabla T=0,\label{eqn:energy0}\\
 &&p=\rho\frac{\calR}{\mu}T,
\end{eqnarray}
where the equation of motion is omitted because of the isobaric
assumption, $\kappa(T)$ is the conductivity, $\calR$ the gas constant,
$\mu$ the mean molecular weight, and $\gamma$ the adiabatic index.  The
heat-loss function, $\calL$, is defined as $\rho\Lambda-\Gamma$, where
$\Lambda$ and $\Gamma$ are the cooling and heating rates, respectively.
Throughout this paper, we assume a single power-law of the conductivity,
$\kappa(T)=\kappa_{0} (T/T_{0})^{\alpha}$, and $\alpha=1/2$ for neutral
gas.  The energy equation (\ref{eqn:energy0}) under the isobaric
condition is derived by applying $T\d s=\d h$, where $\d p/\rho$ is
neglected for the constant pressure, and $s$ and $h$ are specific
entropy and enthalpy, respectively [see, e.g., \citet{pb70} and
\citet{gl73}].  Dividing variables by characteristic values, we obtain
dimensionless quantities, $\tilT=T/T_{0}, \tilrho=\rho/\rho_{0}$, and
$\tilp=p/p_{0}$, and introducing characteristic length $l_{\rm F}$ and
time-scale $t_{0}$, we obtain $\tilde{\bf v}=\bv/(l_{F}/t_{0})$.  Here
we take $t_{0}$ as the cooling time-scale,
\begin{equation}
 t_{0}=\frac{\gamma}{\gamma-1}\frac{\calR T_{0}}{\mu\rho_{0}\Lambda_{0}},
\end{equation}
where $\Lambda_{0}$ is the cooling rate at the characteristic temperature,
$T_{0}$, and $l_{F}$ as the Field length \citep{f65},
\begin{equation}
 l_{F}=\sqrt{\frac{\kappa_{0}T_{0}}{\rho_{0}^{2}\Lambda_{0}}}.
\end{equation}
Using these quantities, the basic equations become the following
dimensionless form,
\begin{eqnarray}
 &&\frac{\partial\tilrho}{\partial\tau}+\nabla\cdot\tilrho\tilde{\bf v}=0,\label{eqn:cont}\\
 &&\frac{\d\tilT}{\d\tau}+\tilL-\frac{1}{\tilrho}\nabla\cdot\tilT^{\alpha}\nabla\tilT=0,\label{eqn:energy}\\
 &&\tilp=\tilrho\tilT,\label{eqn:eos}
\end{eqnarray}
where $\tau=t/t_{0}$ and $\tilL=\calL/\rho_{0}\Lambda_{0}$.  

It is useful to transform these equations into the Lagrangian
description particularly in one dimensional case as shown in ERS92.
Using a relation $\d m=\tilrho \d x$, where $x$ is the dimensionless
Eulerian coordinate normalised by $l_{F}$ and $m$ is the Lagrangian
coordinate, the energy equation (\ref{eqn:energy}) is converted as
follows,
\begin{equation}
 \partial_{\tau}\tilT=-\tilL+\tilp\partial_{m}\tilT^{\alpha-1}\partial_{m}\tilT,
\end{equation}
where $(\tau,m)$ are the independent dimensionless variables and
$\partial_{t}$ and $\partial_{m}$ denote $(\partial/\partial t)$ and
$(\partial/\partial m)$, respectively.  Further transformation
introducing $Z=\tilT^{\alpha}$ makes the above equation simpler,
\begin{equation}
 \partial_{\tau}Z=F(Z)+\tilp Z^{\beta}\partial_{m}^{2}Z^{2},\label{eqn:1dL}
\end{equation}
where $\beta=1-1/\alpha$ and $F(Z)=-\alpha Z^{\beta}\tilL$.  Note that
this simple form is obtained only in the one dimensional case.  ERS92
assumed a simple form of $F(Z)$, which mimics the real heat-loss
function,
\begin{equation}
 F(Z)=Z^{\beta}[\Delta^{2}(Z-Z_{2})-(Z-Z_{2})^{3}-D\log\tilp],\label{eqn:F}
\end{equation}
where $D$ is a positive constant.

To obtain a travelling wave solution, we define $\chi=m-c_{m}\tau$,
where $c_{m}$ is a constant which is a velocity on the Lagrangian
coordinate.  Then the partial differential equation (\ref{eqn:1dL})
becomes an ordinary differential equation,
\begin{equation}
 \tilp Z^{\beta}Z''+c_{m}Z'+F(Z)=0,
\end{equation}
where the prime means differentiation with respect to $\chi$.

An analytic kink solution connecting the two stable phases is obtained
when $c_{m}=0$ as shown in ERS92.  The solution $Z_{0}$ is
\begin{equation}
 Z_{0}(m)=Z_{2}+\Delta\tanh\left[\frac{\Delta}{\sqrt{2}}(m-m_{c})\right].\label{eqn:1dsol}
\end{equation}

Unfortunately, as mentioned above and as shown in \citet{sr94}, it is
difficult to extend it to higher dimensional cases.  This can be easily
seen if we take a $d$-dimensional ($d\geq 2$) continuity, $\d m\propto
r^{d-1}\d x$.  Because this dependence on $d$ produces a term explicitly
depending on $m$, the second derivative $\partial_{m}^{2}$ cannot be
regarded as $\nabla_{m}^{2}$.  To avoid this difficulty, \citet{sr94}
assumed that a time-dependent GL (TDGL) equation \citep[e.g.][]{b94} in
the Lagrangian space is a good approximation for the evolution of a
spherical cloud.  To be free from such an assumption, it is fruitful to
be back in the Eulerian space to treat the fluid equations in higher
dimensions.

\section{Higher dimensional cases: spherical and cylindrical geometry}
In the Eulerian space ($\tau,r$), introducing $X=\tilT^{1+\alpha}$, we
obtain the continuity and energy equations,
\begin{eqnarray}
 && \frac{\d X}{\d\tau}-(1+\alpha)X\nabla\cdot\bv=0,\label{eqn:contE0}\\
 && \frac{\d X}{\d\tau}=\frac{1+\alpha}{\alpha}X^{1/(1+\alpha)}F[Z]+\frac{1}{\tilp}X\nabla^{2}X.\label{eqn:energyE0}
\end{eqnarray} 
Transforming the coordinate to $\chi=r-R_{d}(\tau)$ for a
$d$-dimensional spherical cloud,
\begin{eqnarray}
 -\dot{R}_{d}X'&=&-vX'+(1+\alpha)Xv'+(1+\alpha)\frac{d-1}{r}vX,\label{eqn:contE}\\
 -\dot{R}_{d}X'&=&-vX'+\frac{1+\alpha}{\alpha}X^{1/(1+\alpha)}F[Z]\nonumber\\
 &&\qquad +\frac{1}{\tilp}X''X+\frac{1}{\tilp}\frac{d-1}{r}X'X,\label{eqn:energyE}
\end{eqnarray} 
where $\dot{R}_{d}=\d R_{d}/\d\tau$ and $F$ is given by
eq.(\ref{eqn:F}).  The radius of the cloud, $R_{d}(\tau)$, is hereafter
defined to be the position at $X=X_{2}\equiv Z_{2}^{(1+\alpha)/\alpha}$.
The fluid velocity $v$ is defined in the rest frame of the centre of the
cloud.  By defining $u_{d}\equiv v-\dot{R}_{d}$, which is the fluid
velocity in the rest frame of the front, the above equations become
\begin{eqnarray}
 u_{d}X'&=&(1+\alpha)Xu_{d}'+(1+\alpha)\frac{d-1}{r}(u_{d}+\dot{R}_{d})X,\label{eqn:contE2}\\
 u_{d}X'&=&\frac{1+\alpha}{\alpha}X^{1/(1+\alpha)}F[Z]+\frac{1}{\tilp}X''X+\frac{1}{\tilp}\frac{d-1}{r}X'X.\label{eqn:energyE2}
\end{eqnarray} 
Because we take $v=0$ at the cloud centre, $u_{d}$ must be
$-\dot{R}_{d}$ in the vicinity of the centre.  In the plane-parallel
case $d=1$, using the relationship $\tilT=X^{1/(1+\alpha)}=Z^{1/\alpha}$
and $\tilrho u_{d}=-c_{m}$, eq.(\ref{eqn:energyE}) can be reduced to
eq.(\ref{eqn:energy}), and therefore the kink solution
eq.(\ref{eqn:1dL}) is easily proved to satisfy these equations when
$\dot{R}_{1}=0$.

The above equations are non-linear, so it is hard to find an analytic
solution.  Therefore we simply assume what follows.  One is that the
structure of a solution is very similar to a one-dimensional ($d=1$)
solution.  The second assumption is that the first derivative of the
solution is sharply peaked at the interface, in other words, the width
of the interface is very narrow compared to the scale of the cloud size.
This corresponds to $X'=0$ at $r\neq R_{d}$.  Under these assumptions,
it is reasonable to substitute the first and second terms of the
right-hand side of eq.(\ref{eqn:energyE2}) into $u_{1}X'$.  Thus we
obtain an approximate form,
\begin{equation}
 u_{d}(R)=u_{1}(R)+\frac{1}{\tilp}\frac{d-1}{R_{d}}X_{2},\label{eqn:u-R}
\end{equation}
where $u_{d}(R)$ is the fluid velocity passing the front at $r=R_{d}$ in
the front rest frame.  $X_{2}$ in the last term emerges because we
define the position of the front at which $X=X_{2}$.  This is similar to
an equation discussed by \citet{gl73}.  Note that the above
approximation corresponds to taking the first order in the expansion of
the curvature term, $1/r=1/R+{\cal{O}}(R^{-2})$, where $1/R$ is the
ratio of the Field length $l_{F}$ to the radius of the cloud
\citep{ams95}.  Thus this is valid only for $R\gg 1$.

\begin{figure}
\epsfxsize=0.9\hsize
\epsfbox{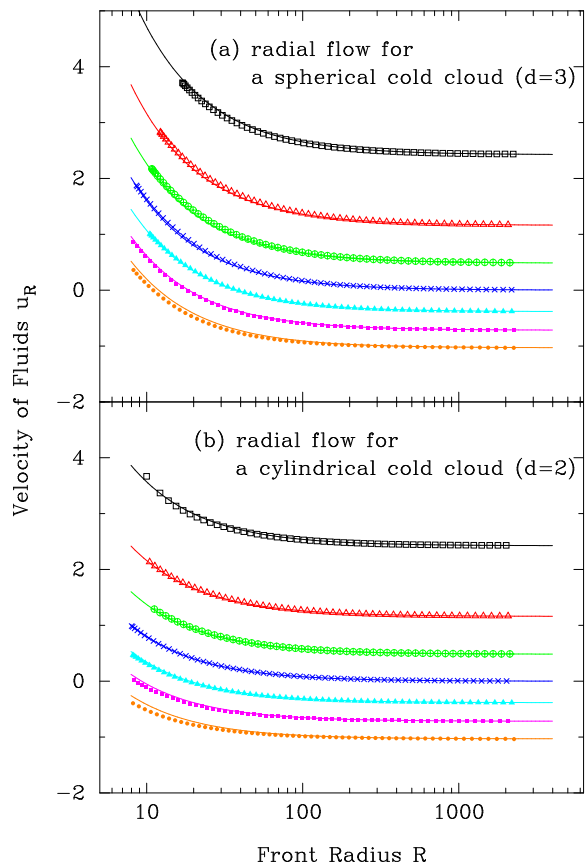}

\caption{Fluid velocity $u_{d}(R)$ against radius $R$ for (a) $d=3$
(sphere) and (b) $d=2$ (cylinder).  The curves and symbols represent the
approximate solution, eq.(\ref{eqn:u-R}), and the numerical solution
given by eqs.(\ref{eqn:contE}) and (\ref{eqn:energyE}).  From the top to
the bottom, $\tilp=0.7, 0.8, 0.9, 1, 1.1, 1.2$ and 1.3, respectively.
We set $Z_{2}=2$ and $\Delta=D=1$ for simplicity.  }

\label{fig:u-R}
\end{figure}

Fig.\ref{fig:u-R} shows the fluid velocity at the front, $u_{d}(R)$,
against $R_{d}$ for $d=2$ and 3.  The curves and symbols represent those
given by the approximate solutions and numerical ones, respectively, for
different values of pressure.  Clearly the above approximate solution
well agrees with the numerical solutions.  In the large limit of
$R_{d}$, we have confirmed that the fluid velocity converges on
$u_{1}(R)$.  Note that the cloud evaporates when $u_{d}$ is positive,
and vice versa.

To estimate the front velocity, we need to know the relationship between
$u_{d}(R)$ and $\dot{R}_{d}$.  Here we evaluate
$f\equiv-\dot{R}_{d}/u_{d}(R)$ from the results for $d=1$ case.  Noting
that $c_{m}$ in eq.(\ref{eqn:1dsol}) corresponds to the mass flux,
$c_{m}=-\tilrho u_{1}$, $u_{1}(R)$ can be written as
\begin{equation}
 u_{1}(R)=-c_{m}X_{2}^{1/(1+\alpha)}/\tilp,
\end{equation}
similarly,
\begin{equation}
 \dot{R}_{1}=-u_{1}(0)=c_{m}X(0)^{1/(1+\alpha)}/\tilp,
\end{equation}
where the argument 0 represents values at $r=0$.  From those,
\begin{equation}
 f=\left[X(0)/X_{2}\right]^{1/(1+\alpha)}.
\end{equation}
Thus, the motion of the front is described by
\begin{equation}
 \dot{R}_{d}=\frac{\d R}{\d\tau}=\dot{R}_{1}-f\frac{1}{\tilp}\frac{d-1}{R_{d}}X_{2}.\label{eqn:AC}
\end{equation}

\begin{figure}
\epsfxsize=0.9\hsize
\epsfbox{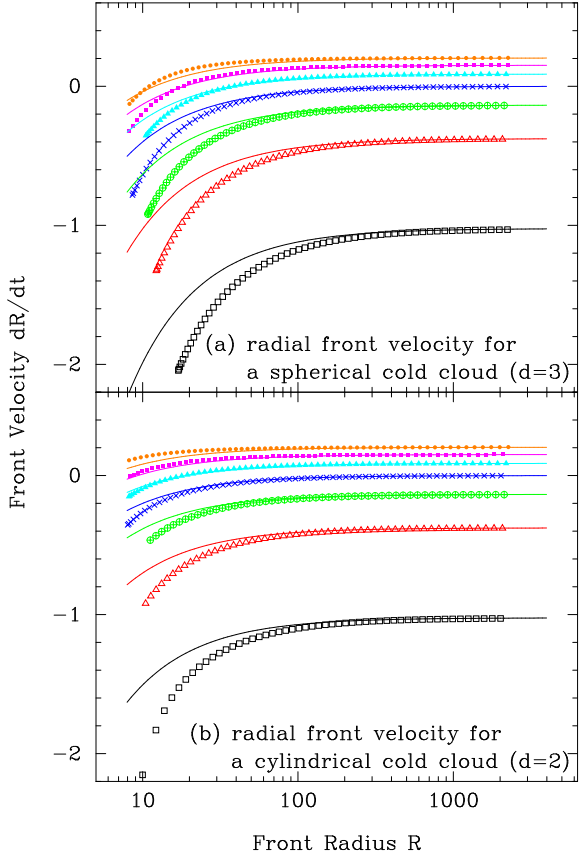}

\caption{The same as Fig.\ref{fig:u-R} but for the front velocity,
$\dot{R}_{d}$.  The curves and symbols represent the approximate
solution, eq.(\ref{eqn:AC}), and the numerical solution given by
eqs.(\ref{eqn:contE}) and (\ref{eqn:energyE}).  From the top to the
bottom, $\tilp=1.3, 1.2, 1.1, 1, 0.9, 0.8$ and 0.7, respectively.  }

\label{fig:v-R}
\end{figure}

Fig.\ref{fig:v-R} shows the same as Fig.\ref{fig:u-R} but for the front
velocity $\dot{R}_{d}$.  At $R_{d}\ga 10^2$, the approximate solutions
well agree with the numerical ones.  On the other hand, At $R_{d}\la
10^2$, the deviation of the approximate solutions from the numerical
ones becomes large.  This might suggest that the way of estimating $f$
is too simple because we have found that the fluid velocity itself is
well approximated by eq.(\ref{eqn:u-R}).  Nevertheless,
eq.(\ref{eqn:AC}) well describes the critical radius, $R_{{\rm crit},d}$,
at which $\dot{R}_{d}=0$,
\begin{equation}
 R_{{\rm
 crit},d}=\frac{d-1}{c_{m}}X_{2}^{\alpha/(1+\alpha)}=-\frac{d-1}{\tilrho
 u_{1}}\tilT_{2}^{\alpha},
\end{equation}
where $T_{2}\equiv X_{2}^{1/(1+\alpha)}$.  In fact, we have found that
we can obtain a better fit when we replace $f$ by $f/\tilp^{1+\alpha}$,
while it makes the good estimation of $R_{{\rm crit},d}$ worse.

In the expression of eq.(\ref{eqn:AC}), an undesirable dependence on the
temperature at the front remains, whose definition also has an
ambiguity.  This should be removed by replacing $X_{2}$ and $R$ in the
curvature term by values at the maximum of $X'X$.  To do this, however,
the thickness of the front should be explicitly considered.  This will
be done in a subsequent work.  Nevertheless, we have found that it
provides a good fit to numerically obtained solutions when the front is
defined at the unstable equilibrium in the case of the saturation
pressure, $\tilp=1$.  Apart from this, we would like to stress that the
above approximate solution can be derived from values $\dot{R}_{1}$ and
$f$ in $d=1$ case.

\begin{figure}
\epsfxsize=0.9\hsize
\epsfbox{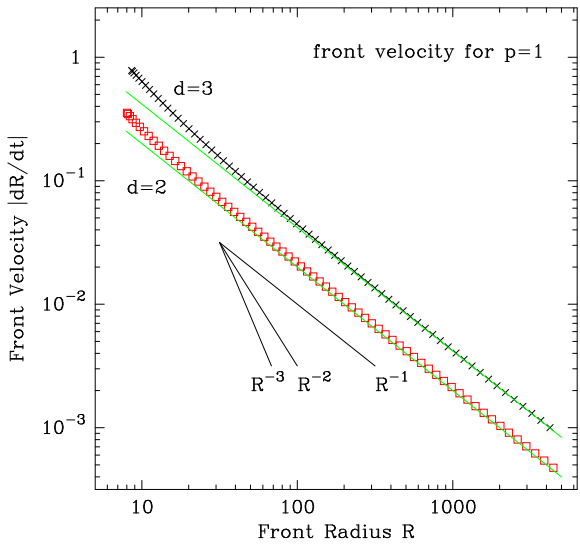}

\caption{The absolute value of the front velocity $|\dot{R}|$ against
$R$.  To see its radial dependence, the vertical axis is logarithmic.
The straight lines and symbols represent the front velocity provided by
the approximate solution and the numerical solution, respectively.  The
upper and lower sets indicate $d=3$ and 2, respectively.  It is evident
that our results are proportional to $R^{-1}$ different from the
prediction by \citet{sr94}.  }

\label{fig:v-R2}
\end{figure}

Finally, we discuss the dependence of the front velocity on the front
position.  As mentioned in Section 1, \citet{sr94} considered the
frontal motion based on a model equation similar to the TDGL equation.
They found that $\dot{R}\propto R^{-d}$ when pressure is nearly equal to
the saturation pressure, and $\dot{R}\propto R^{-(d-1)}$ when far from
the saturation pressure.  Because we have already shown in
Fig.\ref{fig:v-R} that $\dot{R}\to $ const. as $R\to\infty$ when far
from the saturation pressure, we show the case of $\tilp=1$, in which we
can see only the curvature effects.  Fig.\ref{fig:v-R2} shows a log-log
plot for $|\dot{R}_{d}|$ against $R$ for $d=2$ and 3.  It is evident
that the front velocity is proportional to $R^{-1}$ independent of $d$.
Thus we conclude that the TDGL equation provided in the Lagrangian
coordinate does not provide a correct model for thermally bistable
fluids.

\section{General curved fronts}
\begin{figure}
\epsfxsize=\hsize
\epsfbox{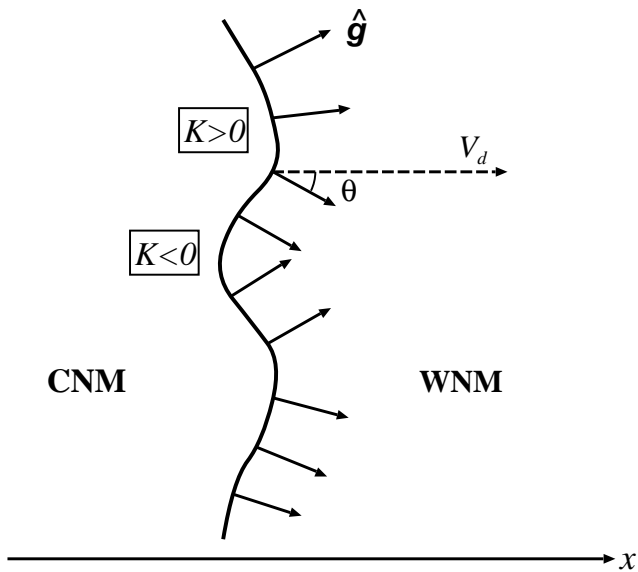}

\caption{ Schematic view of a curved front.  The thick wavy curve
represents the front connecting the CNM ({\it left}) and WNM ({\it
right}).  The solid arrows denote the normal vector $\ghat$, and the
dashed arrow the velocity of the front along the horizontal axis
$V_{d}$.  }

\label{fig:geo}
\end{figure}

Using the curvature term derived in the previous section, we discuss the
dynamics of general curved fronts according to \citet{b94}.  We define
the $x$-axis normal to the unperturbed (straight) front, and take a unit
vector normal to the curved front, $\ghat$, with a direction from the
CNM to the WNM.  The situation under consideration is shown in
Fig.\ref{fig:geo} schematically.  Using this, we can write $\nabla
X=(\partial_{g}X)_{\tau}\ghat$ and
$\nabla^{2}X=(\partial_{g}^{2}X)_{\tau}+(\partial_{g}X)_{\tau}\nabla\cdot\ghat$.
Substituting these into equation (\ref{eqn:energyE2}), we obtain
\begin{eqnarray}
 u_{d}\left(\partial_{g}X\right)_{\tau}
  &=&\left[\left(\partial_{g}^{2}X\right)_{\tau}
   +\left(\partial_{g}X\right)_{\tau}\nabla\cdot\ghat\right]\frac{X}{\tilp}\nonumber\\
  &&+\frac{1+\alpha}{\alpha}X^{1/(1+\alpha)}F[Z],
\end{eqnarray}
As done in the previous section, substituting corresponding terms in the
right hand side with the one-dimensional equation and noting the
existence of the direction cosine with respect to the normal direction
of the unperturbed front, we obtain
\begin{equation}
 V_{d}=\cos\theta[\dot{R}_{1}-fX_{2}K/\tilp],
\end{equation}
where $V_{d}$ is the velocity of the front along the $x$-axis, and
$K=\nabla\cdot\ghat$ is the mean curvature.  When we take $R_{d}$ as a
curvature radius, we obtain $K=(d-1)/R_{d}$.  This result is applicable
to general curved fronts.

Let us consider a convex region of the CNM against the WNM, in which
$K>0$.  When the CNM is evaporating ($\tilp<1$), the sign of
$\dot{R}_{1}$ is negative.  Because the sign of the second term is
always negative, the CNM in the convex region evaporates faster than in
concave regions.  When the CNM is accreting cooling WNM ($\tilp>1$), the
sign of $\dot{R}_{1}$ is positive.  Thus the condensation in the convex
region is slower then in concave regions.  Consequently, the curvature
term always smooths out the curved front.  Note that this conclusion is
valid only under the assumptions that flows are normal to the front and
that the structure of the front is the same as that in the case of
plane-parallel geometry, in addition to the approximation of isobaric
evolution.

\section{Conclusions and Discussion}
We have investigated the dynamics of thermally bistable fluids from a
pattern-theoretical point of view.  To evaluate the curvature effects of
the front connecting the WNM and CNM, at first, we focused on
$d$-dimensional spherically symmetric CNM clouds.  Using a way of
approximations often used in the field of pattern formation theories, we
derived an approximate solution describing the velocity of the front.
Comparing with numerical solutions, we confirmed that they are in good
agreement with each other, at least when the cloud size is larger than a
few tens times the Field length, which corresponds to the thickness of
the front.  We have also found that our results contradict those given
by \citet{sr94}, which assumed a model equation similar to the TDGL
equation.  We showed that the velocity of the front is proportional to
the inverse of the radius in the case of the saturation pressure, and is
constant in the case of much larger clouds and/or pressure far from the
saturation pressure.  Second, we discussed the dynamics of general
curved fronts.  Using the obtained approximate solution, we have found
that the curvature effects smooth out curved fronts.

In contrast to the latter conclusion, recent numerical experiments of
interstellar turbulence \citep{ki02b, ki05} show that most fronts might
be unstable even without strong shocks.  This apparent contradiction
might come from our simple assumptions that the structure of the front
is independent of geometry, that the fluid motion is normal to the
front, and that the fluid evolves isobarically.  Presumably some
instability mechanisms are there similar to the Darrieus-Landau
instability in propagating flame \citep{ll87, iik05}, or the
Mullins-Sekerka instability in crystal growth \citep{ms63, ms64}.  This
work is a first step toward full understanding of the dynamical
behaviour of the ISM found in the numerical experiments.  For fair
comparison, we must clarify at least how the above assumptions affect
the dynamics of the ISM.  We are planning to analyse the stability of
the front considering these known mechanisms, as well as the validity of
the isobaric evolution.  Together with those, hydrodynamic simulations
should be performed to be compared with the results obtained in this
paper.

\section*{ACKNOWLEDGMENTS}    

We would like to thank Tsuyoshi Inoue for useful discussion.  MN is
supported by the Japan Society for the Promotion of Science for Young
Scientists (No.207).  MN also thank Yasuhiro Hieida for inviting me to
the world of non-equilibrium statistical physics.  Si is supported by
the Grant-in-Aid (No.15740118, 16077202) from the Ministry of Education,
Culture, Sports, Science, and Technology (MEXT) of Japan.

\bsp

\end{document}